\documentclass[a4paper,final]{appolb}
\usepackage{amsfonts,amsmath,amssymb,bm}
\usepackage{graphicx} 
\usepackage{pstricks}
\usepackage{subfig}

\newcommand{\be}{\begin{equation}}
\newcommand{\bea}{\begin{eqnarray}}
\newcommand{\ee}{\end{equation}}
\newcommand{\eea}{\end{eqnarray}}

\def\s#1{{\scriptscriptstyle #1}}
\def\n#1{({\it #1}\,)}

\def\noeq#1{(\ref{#1})}
\def\1eq#1{Eq.~(\ref{#1})}
\def\2eqs#1#2{Eqs.~(\ref{#1}) and~(\ref{#2})}
\def\3eqs#1#2#3{Eqs.~(\ref{#1}), (\ref{#2}) and~(\ref{#3})}
\def\4eqs#1#2#3#4{Eqs.~(\ref{#1}), (\ref{#2}), (\ref{#3}) and~(\ref{#4})}

\def\fig#1{Fig.~\ref{#1}}
\def\ie{{\it i.e.}, }
\def\eg{{\it e.g.}, }

\begin{document}

\title{Recent results in the infrared sector of QCD}

\author{D. Binosi\thanks{Talk presented at the International Meeting ``Excited QCD'', Peniche, Portugal, 6 - 12 May, 2012}
\address{European Centre for Theoretical Studies in Nuclear
Physics and Related Areas (ECT*)\\ and\\ Fondazione Bruno Kessler,\\ Villa Tambosi, Strada delle
Tabarelle 286, 
I-38123 Villazzano (TN)  Italy}}

\maketitle

\begin{abstract}
\noindent We review the most recent results, derived within the combined framework of the pinch technique and the background field method, describing certain QCD nonperturbative properties. 
\end{abstract}

Since its introduction, the combined framework of the pinch technique (PT)~\cite{Cornwall:1981zr,Cornwall:1989gv,Binosi:2002ft,Binosi:2003rr,Binosi:2009qm} and the background field method~\cite{Abbott:1980hw}, known in the literature as the PT-BFM scheme~\cite{Aguilar:2006gr,Binosi:2007pi,Binosi:2008qk}, has provided a sound theoretical basis for addressing the nonperturbative study of the QCD Green's functions of both the gluon as well as the ghost sector of QCD, respecting at the same time the fundamental symmetries of the theory. Results derived within this framework include, but are not limited to, the first evidence of the existence (in the Landau gauge) of massive solutions in the properly truncated QCD Schwinger-Dyson equations (SDEs) as found in lattice simulations~\cite{Aguilar:2008xm} -- and which can be interpreted  in terms of  a nonperturbative mass~\cite{Aguilar:2008xm,RodriguezQuintero:2010ss,Pennington:2011xs}
which tames the infrared (IR) divergences of the Green's functions of the theory--,  the study of the Kugo-Ojima function~\cite{Aguilar:2009pp} and the identification of the role of the ghost for achieving a chiral symmetry breaking pattern that provides for dynamically generated quark masses compatible with phenomenology~\cite{Aguilar:2010cn}. 

In this talk I will present the latest results derived within the PT-BFM framework (in the Landau gauge) and discuss in particular:

\begin{itemize}
\item The use of the SDEs to  compute   the
nonperturbative  modifications caused  to the  IR  finite gluon
propagator by the inclusion of a small number of
quark families~\cite{Aguilar:2012rz};
\item The  general  derivation  of  the  full  non-perturbative
equation that governs the momentum evolution of the  dynamically generated gluon mass~\cite{Binosi:2012sj}. 
\end{itemize}

\section{Unquenching the gluon propagator}

As described in~\cite{Aguilar:2012rz} the PT-BFM allows to develop an approximate  method for  ``unquenching'' the  (IR  finite) gluon propagator,  computing nonperturbatively  the
effects induced by a small number of light quark families.
The procedure consists of two basic steps:
\begin{itemize}

\item Computing the fully-dressed quark-loop diagram, using as input the nonperturbative quark propagators 
obtained from the solution of the gap equation, together with an Ansatz for the fully-dressed quark-gluon vertex that preserves gauge-invariance~\cite{Aguilar:2010cn};

\item Adding this result  to the quenched gluon propagator obtained in large-volume lattice simulations.  

\end{itemize}

The key assumption of the method sketched above, is therefore that the effects of a small number of quark families to the gluon propagator may be considered as a ``perturbation'' to the quenched case, of which the quark-loop diagram constitutes the leading correction term, with the subleading terms stemming   
from the (originally) pure Yang-Mills diagrams which now get modified from the quark loops nested inside them.
Thus, within the approximations we will employ these latter corrections are neglected, so that one can identify  (even when dynamical quarks are present) with the quenched lattice propagator all SDE graphs except the quark loop diagram. 

The expression for the PT-BFM scalar cofactor $\Delta_\s Q(q^2)$ of the unquenched propagator  (the subindex ``Q'' standing for ``quarks''), defined as\linebreak \mbox{$\Delta^{\mu\nu}_\s Q(q)=P_{\mu\nu}(q)\Delta_\s Q(q^2)$} with $P_{\mu\nu}=g_{\mu\nu}-q_\mu q_\nu/q^2$ the dimensionless transverse projector,  can be then written as~\cite{Aguilar:2012rz}
\be
\Delta_{\s{Q}}(q^2) = \frac{\Delta(q^2)}
{1 + \left\{ i \,\widehat{X}(q^2) \left[1+G(q^2)\right]^{-2}- \lambda^2 \right\}\Delta(q^2)}.
\ee
In what follows we will describe all the different terms appearing in the right-hand side of the formula above.

\begin{itemize}

\item  $\Delta(q^2)$ is the quenched propagator which, as already pointed out, will be identified with the one obtained from the large volume lattice simulations.

\item $G(q^2)$ is a special Green's function particular to the PT-BFM which achieves the conversion from the PT-BFM to the conventional gluon propagator~\cite{Binosi:2007pi,Binosi:2008qk}; in the Landau gauge it is known to coincide with the Kugo-Ojima function~\cite{Aguilar:2009pp,Grassi:2004yq}.

\item $\widehat{X}(q^2)$ is the PT-BFM scalar cofactor resulting from the calculation of the quark loop diagram; defining  $\widehat{X}_{\mu\nu}(q)=P_{\mu\nu}(q)\widehat{X}(q^2)$, one has 
\be
\widehat{X}(q^2)=-\frac{g^2}{12}\!\int_k\mathrm{Tr}\left[
\gamma^\mu S(k)\widehat\Gamma_\mu(k+q,-k,-q)S(k+q)\right],
\label{qse}
\ee
where $S$ is the full fermion propagator (with $S^{-1}(p)=-iA(p)[p\hspace{-0.17cm}/-{\cal M}(p)]$ and ${\cal M}$ the dynamical quark mass), and $\widehat\Gamma_\nu$ is the full PT-BFM quark-gluon vertex. To evaluate expression~\noeq{qse} one proceeds as follows: \n{i} the  nonperturbative behavior of  the functions $A$
and  ${\cal M}$   appearing  in  the   definition  of  the   full  quark
propagator are obtained by solving numerically the quark gap equation as done in~\cite{Aguilar:2010cn} using a vertex Ansatz improved with the inclusion of the (numerically crucial)  dependence on the ghost dressing function and the quark-ghost scattering amplitude~\cite{Aguilar:2010cn}; \n{ii} for the full PT-BFM quark-gluon vertex $\widehat{\Gamma}$  one uses a suitable nonperturbative Ansatz,
satisfying the gauge symmetry of the theory --such as the Ball-Chiu vertex~\cite{Ball:1980ay} or the Curtis-Pennington vertex~\cite{Curtis:1990zs}. To be sure,  other forms of the quark-gluon vertex exists, such as those reported  in~\cite{Kizilersu:2009kg,Bashir:2011dp}, and it would be interesting to check what effects they might have on our predictions.

\item Finally, $\lambda^2=\Delta^{-1}_\s{Q}(0)-\Delta^{-1}(0) \equiv m^2_{\s{Q}}(0) - m^2(0)$ denotes the gluon mass difference at $q^2 =0$ (notice that since $\widehat{X}(0)=0$, the quark contribution to this quantity is only indirect,\ie through the modification it will induce on the various ingredients appearing in the mass equation -- see next section). A solid first-principle determination of $\lambda^2$  
has not been attempted, mainly due  to the 
fact that the derivation of the {\it complete} mass equation has been only very recently achieved~\cite{Binosi:2012sj} (see the next section again); in the analysis presented here we will restrict ourselves to extracting  an approximate range for $\lambda^2$, by employing  a suitable extrapolation of the (unquenched) curves obtained from intermediate momenta towards the deep IR.

\end{itemize}

\begin{figure}[!t]
\hspace{-.5cm}
\begin{minipage}[b]{0.45\linewidth}
\centering
\includegraphics[scale=0.4]{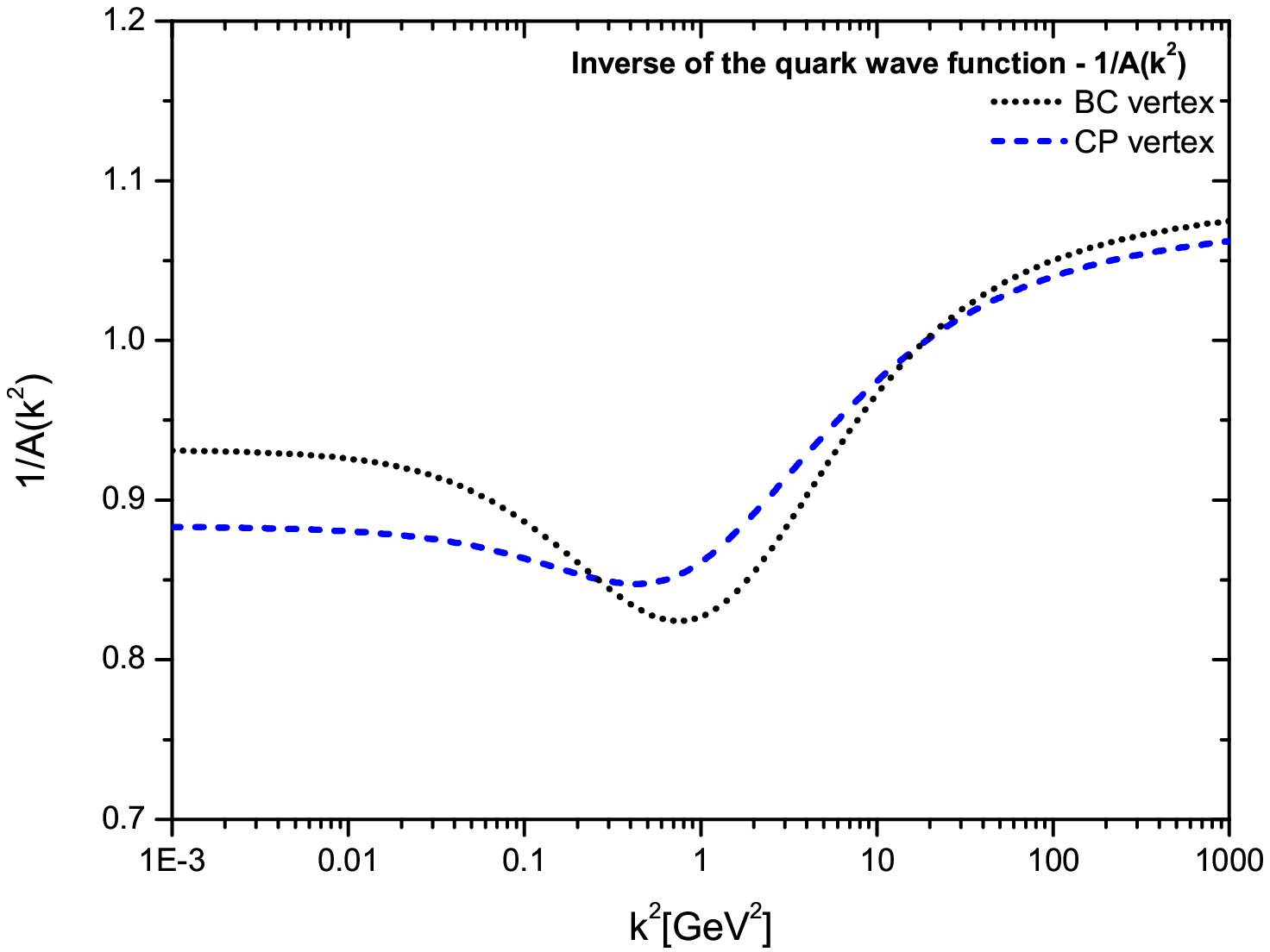}
\end{minipage}
\hspace{0.5cm}
\begin{minipage}[b]{0.50\linewidth}
\includegraphics[scale=0.41]{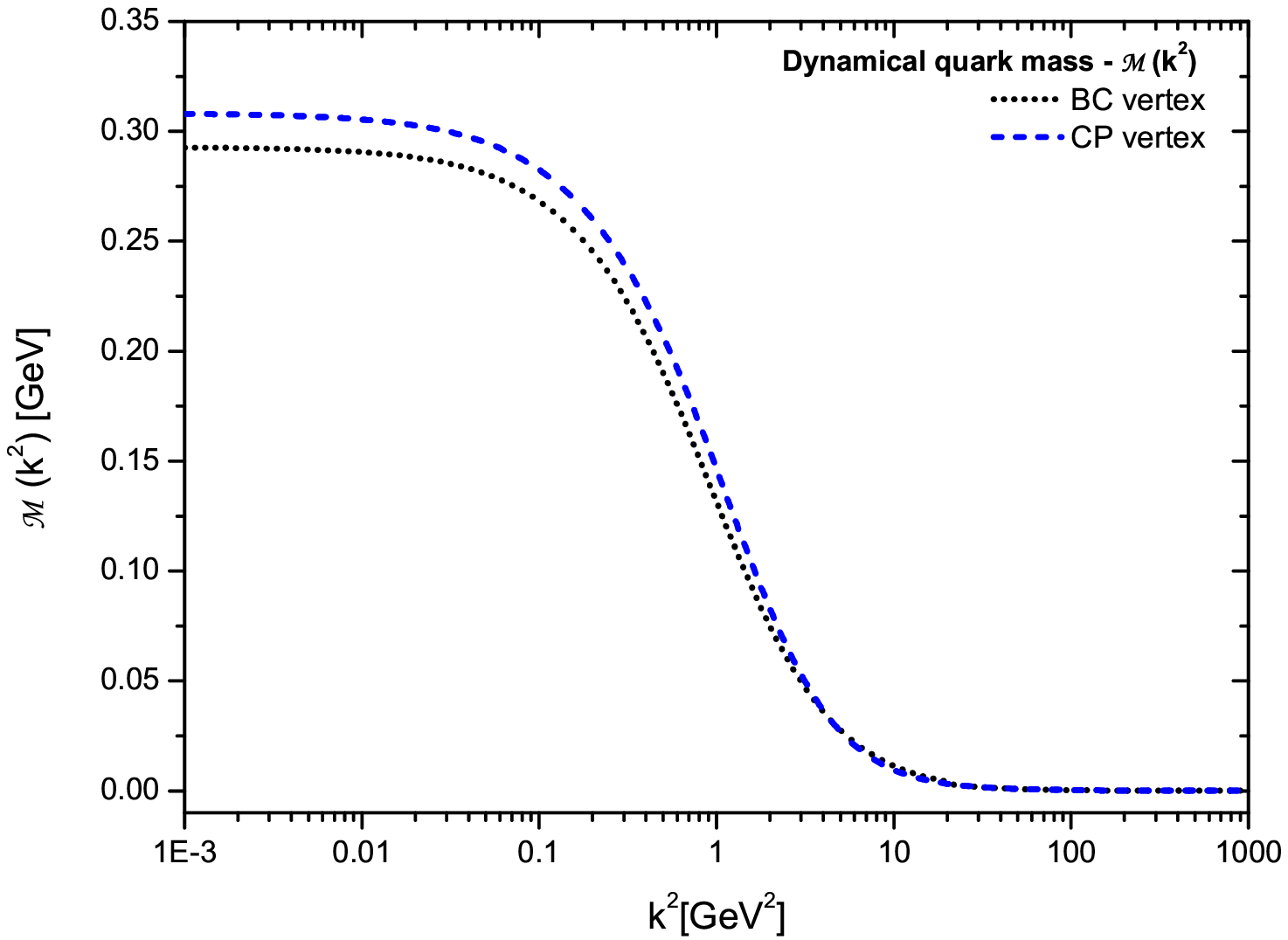}
\end{minipage}
\begin{minipage}[b]{0.45\linewidth}
\centering
\hspace{-.1cm}
\includegraphics[scale=0.385]{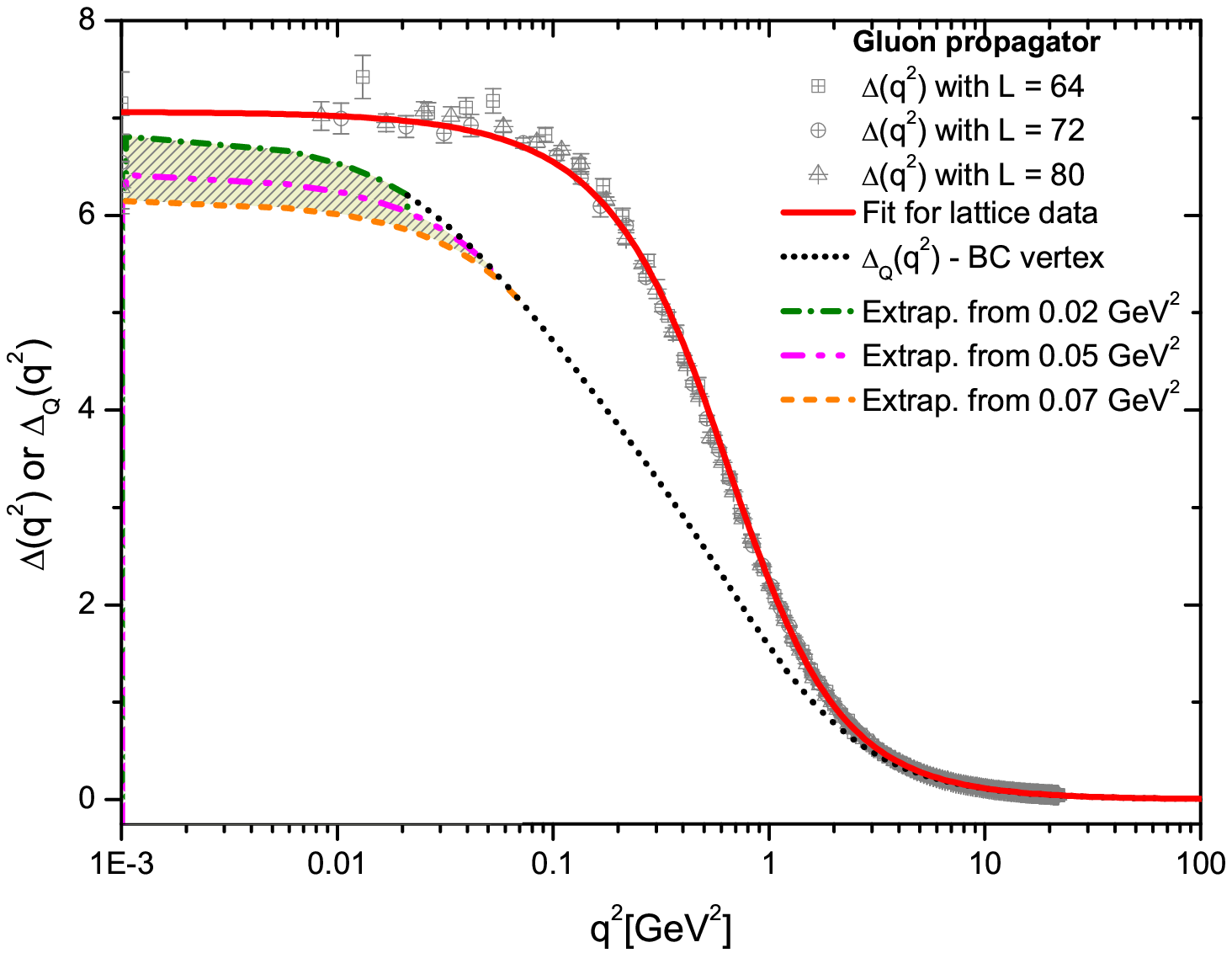}
\end{minipage}
\hspace{0.7cm}
\begin{minipage}[b]{0.50\linewidth}
\includegraphics[scale=0.39]{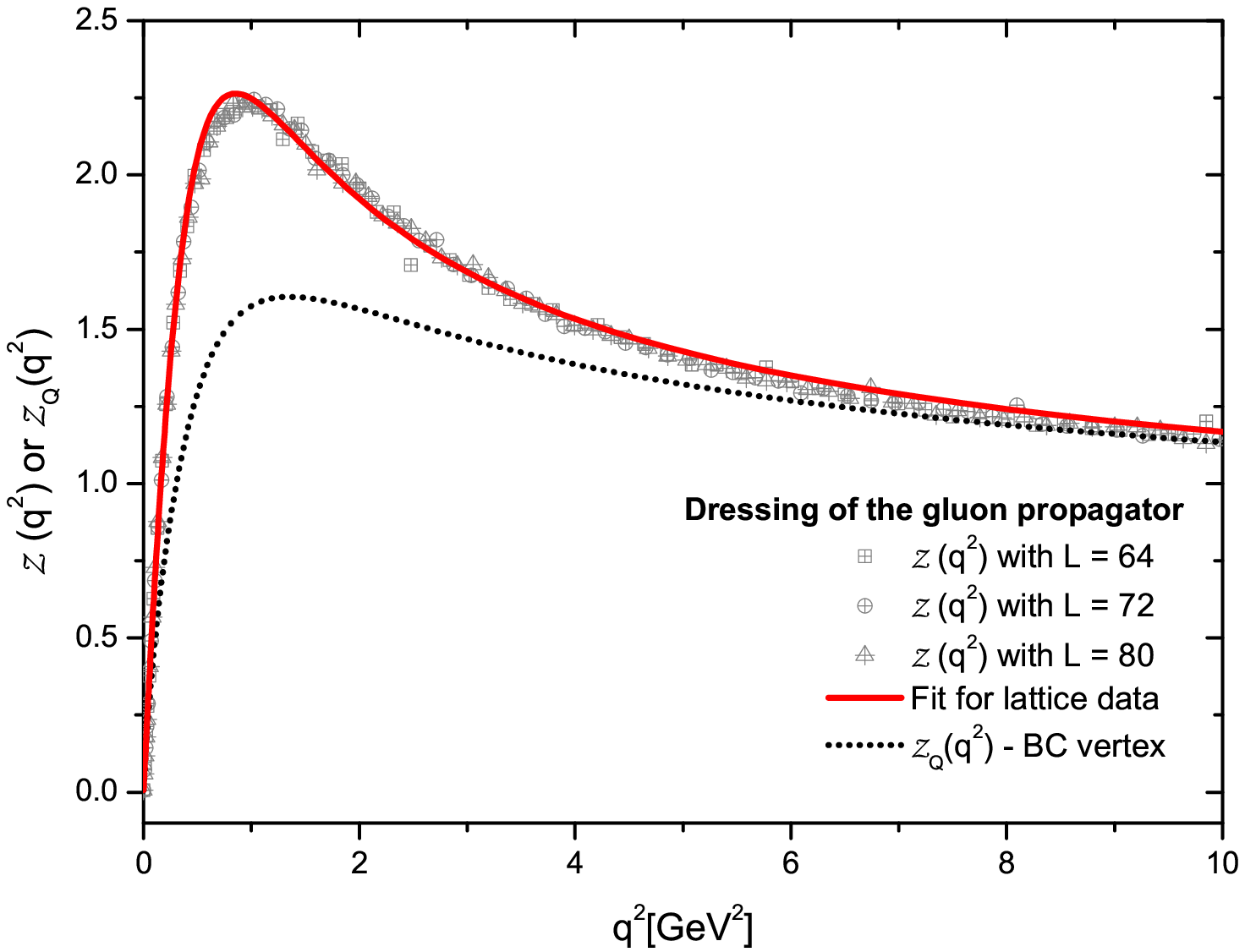}
\end{minipage}
\caption{\label{fig:unquenching1}({\it Top panels}) Solution of the quark gap equation:  $A^{-1}(k^2)$ (left) and dynamical quark mass \mbox{${\mathcal M}(k^2)$} (right) renormalized at $\mu= 4.3$ GeV; dotted black curves correspond to the  Ball-Chiu vertex, while dashed blue curves to the  CP vertex. ({\it Bottom panels}) Comparison between the quenched and the unquenched gluon propagator (left) and dressing function (right). The shaded striped band in the left plot shows
the possible values that $\Delta_{\s Q}(0)$ can assume at zero momentum depending on the extrapolation point used; in the case of the dressing function (which is basically insensible to the IR saturation point) we used a curve with an extrapolation point at $q^2=0.05\,\mbox{GeV}^2$. The quenched lattice results of~\cite{Bogolubsky:2009dc} are also displayed for comparison.}
\end{figure}

\begin{figure}
\begin{minipage}[b]{0.45\linewidth}
\centering
\hspace{-.1cm}
\includegraphics[scale=0.378]{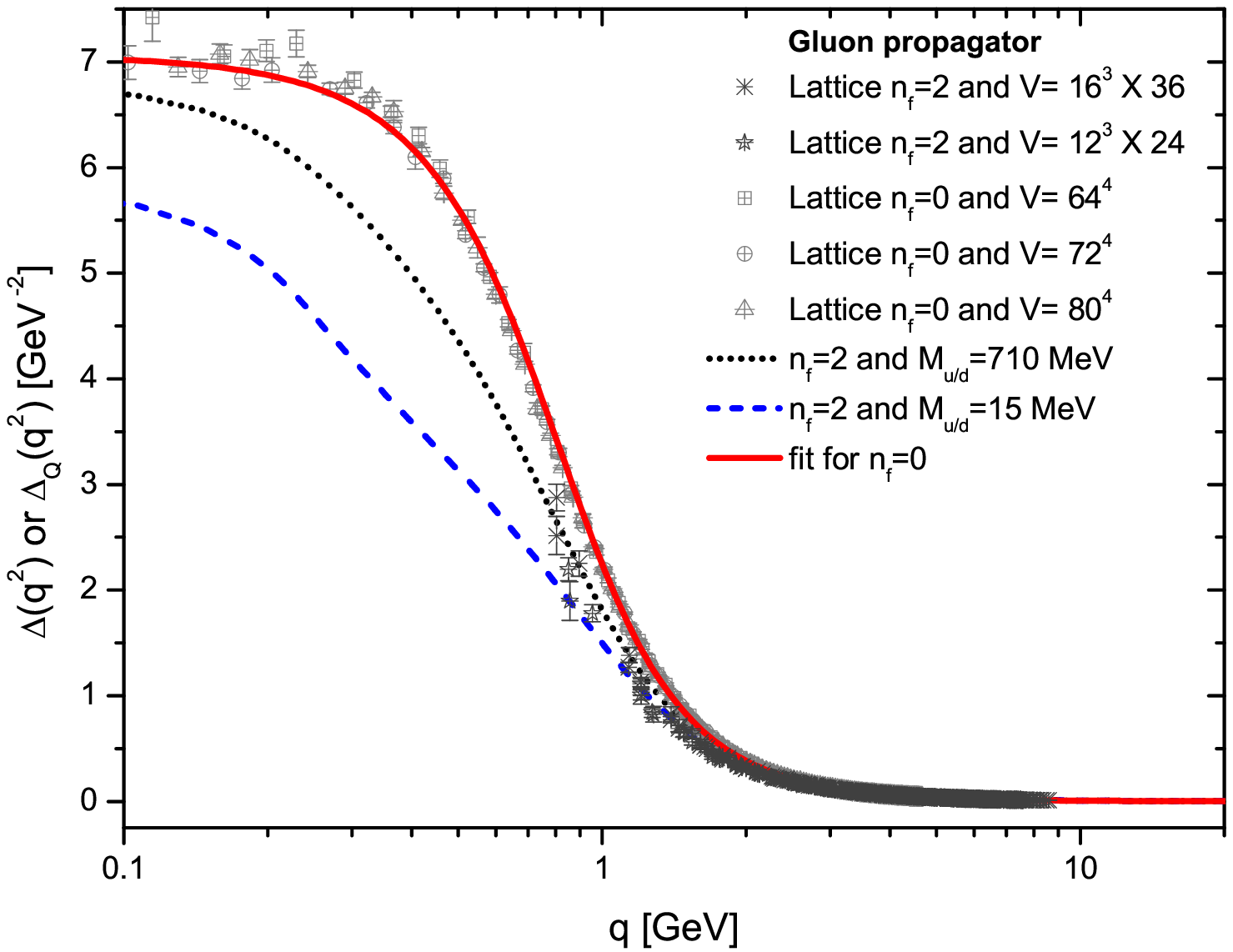}
\end{minipage}
\hspace{0.7cm}
\begin{minipage}[b]{0.50\linewidth}
\includegraphics[scale=0.39]{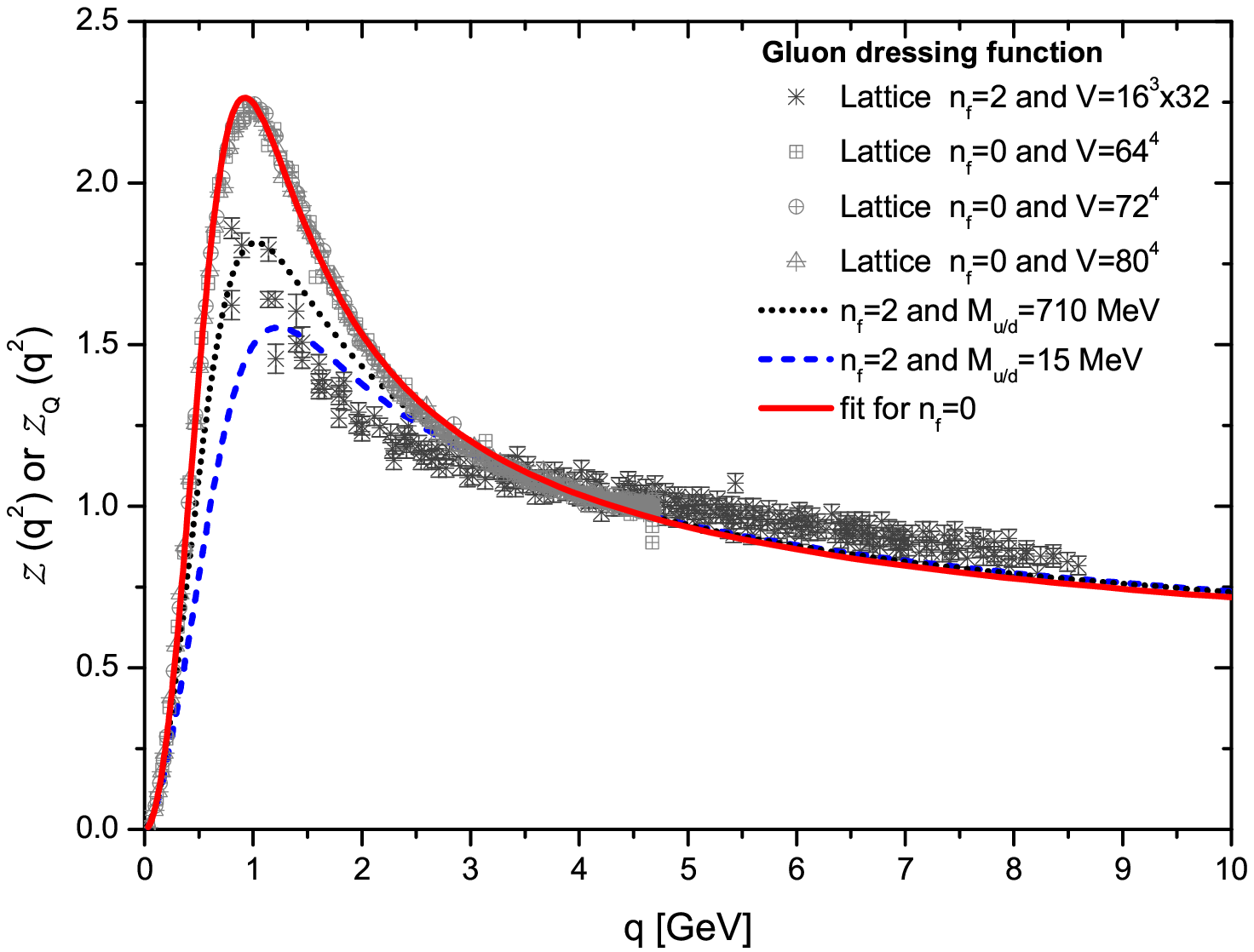}
\end{minipage}
\caption{\label{fig:unquenching2}The unquenched gluon propagator (left) and dressing function (right) obtained in~\cite{Kamleh:2007ud} (dark gray stars), together with the SDE results for two light quarks with \mbox{$M_{u/d}=15$ MeV} (dashed blue curve ) and   \mbox{$M_{u/d}=710$ MeV} (dotted black curve). Quenched data~\cite{Bogolubsky:2009dc} are again shown for comparison.}
\end{figure}

The main results of our study may be summarized as follows~(see \fig{fig:unquenching1}).
The basic effect of the quark loop(s) (one or two families with a constituent mass of the order of 300 MeV)  
is to suppress considerably the gluon propagator in the IR and intermediate momenta regions, while the ultraviolet tails increase, exactly as expected from the standard renormalization group analysis. 
In addition, the inclusion of light quarks makes the gluon propagator saturate at a lower point,  which can be translated into having a larger gluon mass. As far as the gluon dressing function ${\cal Z}(q^2)=q^2\Delta(q^2)$ is concerned, one observes a suppression of the intermediate momentum region peak. A comparison with the recent full QCD lattice simulations of~\cite{Ayala:2012pb} is currently underway;  however a  comparison with some of the available lattice data~\cite{Kamleh:2007ud} (\fig{fig:unquenching2}) shows an excellent qualitative agreement as well as a rather favorable quantitative agreement (with discrepancies at the 20\% level maximum). 

\section{The complete gluon mass equation}

Massive solutions of the gluon propagator SDE can be parametrized as (Euclidean space) $\Delta^{-1}(q^2)=q^2J(q^2)+m^2(q^2)$; therefore one faces the fundamental question of how to disentangle from the SDE the part that determines the evolution of the mass $m^2(q^2)$
from the part that controls the evolution of the ``kinetic'' term $J(q^2)$. This is to be contrasted to what happens in
the analogous studies of chiral symmetry breaking,  
where one derives a system of two coupled equations, 
one determining the ``wave function'' (``kinetic part'')
of the quark self-energy, and one determining the 
dynamical (constituent) quark mass~\cite{Aguilar:2010cn,Roberts:1994dr}. 
Of course, in the 
case of the quark self-energy the above separation 
of both sides of the corresponding SDE (quark gap equation)
is realized in a direct way,  
due to the distinct 
Dirac properties of the two quantities appearing in it, while 
in the case of the gluon propagator no such straightforward  separation is possible. However, 
an unambiguous way for implementing this separation, which exploited to the fullest  
the characteristic structure of a certain type of vertices that are inextricably connected with the 
process of gluon mass generation and naturally appears in the PT-BFM framework, was recently presented in~\cite{Binosi:2012sj}.    

Specifically, a crucial condition for obtaining out of the SDEs 
an IR-finite gluon propagator without interfering with the 
gauge invariance of the theory, is the existence of a set of special vertices that are purely longitudinal and contain massless poles, and must be added to the usual (fully-dressed) vertices of the theory.
The role of these vertices is two-fold. On the 
one hand, thanks to the massless poles they contain, they make possible the emergence of a 
IR finite solution out of the SDE governing the gluon propagator; this corresponds essentially to a non-Abelian realization of the well-known Schwinger mechanism~\cite{Schwinger:1962tn,Schwinger:1962tp}.  
On the other hand, these same poles act like composite Nambu-Goldstone  excitations,  
preserving the form of the STIs of the theory 
in the presence of a gluon mass.

It turns out that the very nature of these vertices furnishes a solid guiding principle for implementing the 
aforementioned separation between mass and kinetic terms. In particular, 
their longitudinal structure, coupled to the fact that one works in the Landau gauge, 
completely determines the longitudinal component of the mass equation; this is tantamount to 
knowing the full mass equation, given that the answer is bound to be transverse. 

Due to the complexity of the derivation of the equation, we will not discuss it here, but rather sketch its final form as well as its main ingredients, together with the numerical solutions it gives rise to. 

Schematically the equation reads
\be
m^2(q^2)=\alpha_s\int_km^2(k^2)\left[{\cal K}_1(\Delta;q,k)+\alpha_s{\cal K}_2(\Delta,Y;q,k)\right],
\label{me-gen}
\ee
where ${\cal K}_1$ is the contribution coming form the one-loop dressed diagrams (namely the graphs appearing in the PT-BFM gluon propagator SDE containing trilinear vertices {\it only}), whereas ${\cal K}_2$ is the contribution of two-loop dressed diagrams (that is, the graphs containing quadrilinear vertices). As indicated in~\1eq{me-gen}, while ${\cal K}_1$ contains only the gluon propagator, in ${\cal K}_2$ a new form factor $Y$ appears, which involves the three gluon vertex and reads
\be
Y(k^2)=\frac1{3k^2}\,k_\alpha g_\beta^\delta \int_\ell\!\Delta^{\alpha\rho}(\ell)\Delta^{\beta\sigma}(\ell+k)\Gamma_{\sigma\rho\delta}(-\ell-k,\ell,k).
\label{defY}
\ee
The lowest order perturbative calculation of  $Y$ (obtained by substituting tree-level values for all quantities appearing in the expression above) yields (after renormalization) $Y\sim\log k^2/\mu^2$; this value multiplied by a constant $C$ (basically modeling, in a rather heuristic way,   
further corrections that may be added to the ``skeleton'' provided by the lowest order result) is the one used in~\cite{Binosi:2012sj} for studying numerically the solutions of \1eq{me-gen}. 
The value of $C$ corresponding to the lowest order expression is fixed  to the actual value $C=3\pi C_A\alpha_s$;
however, it is convenient to treat $C$ as a free parameter, thus
disentangling it from the value of $\alpha_s$, 
and studying what happens to the solution spectrum of \1eq{me-gen} when the two parameters are varied independently.

\begin{figure}
\begin{minipage}[b]{0.45\linewidth}
\centering
\hspace{-.3cm}
\includegraphics[scale=0.5]{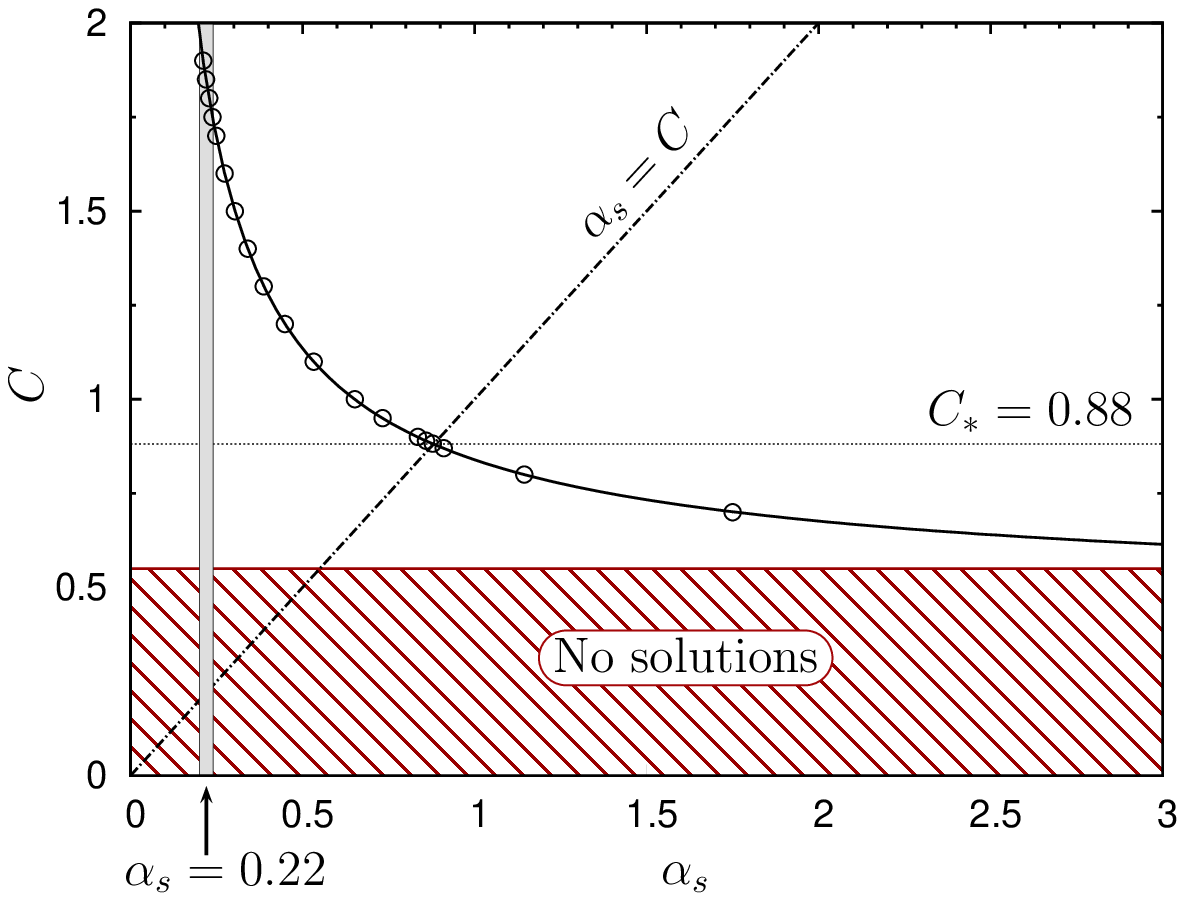}
\end{minipage}
\begin{minipage}[b]{0.50\linewidth}
\includegraphics[scale=0.495]{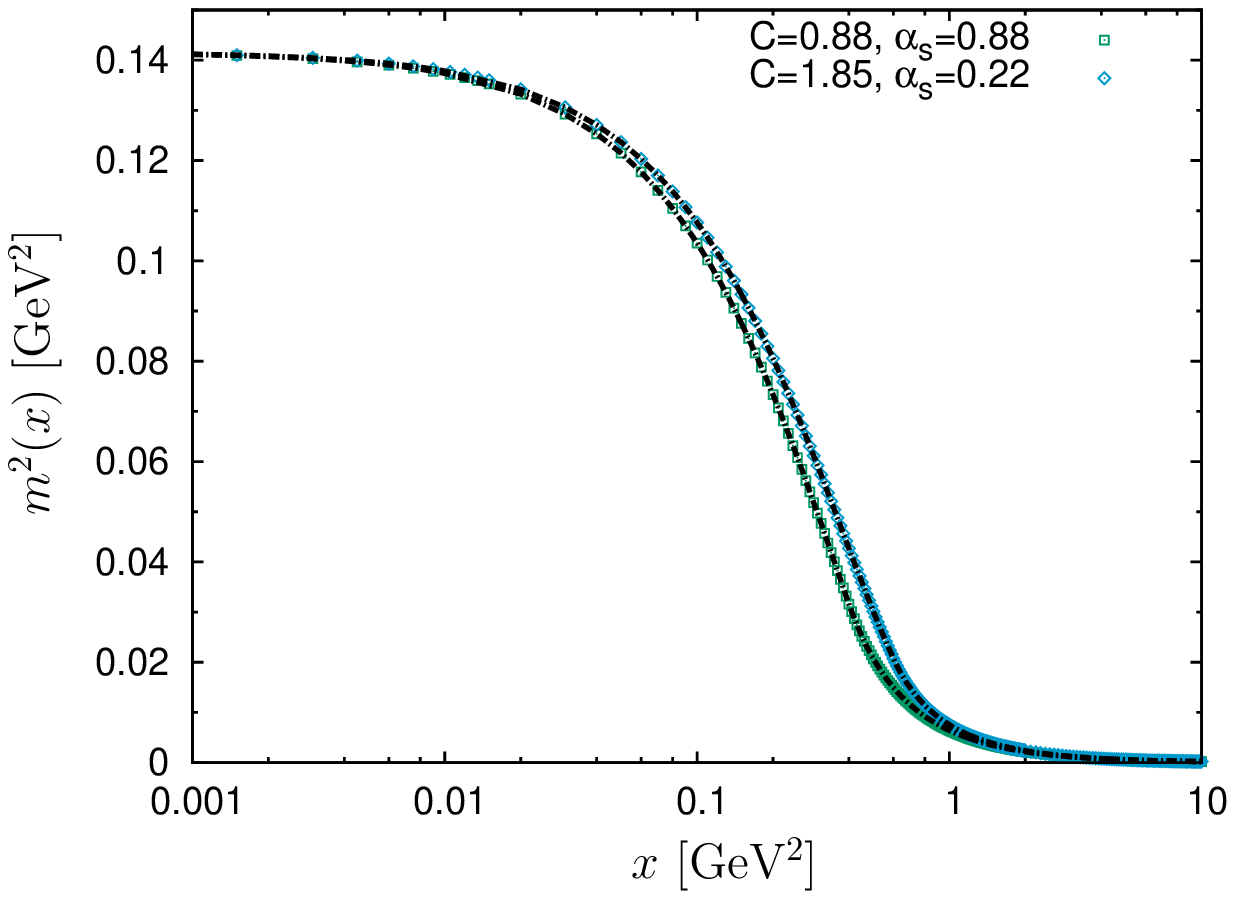}
\end{minipage}
\caption{\label{masseq}The curve described by the set of the pairs $(C,\alpha_s)$ for which one finds physical solutions to the full mass equation~\noeq{me-gen} (left), and the typical monotonically decreasing solution of the mass equation~\noeq{me-gen} (right).}
\end{figure}

As shown in the left panel of~\fig{masseq} (where $C$ is now measured in units of $3\pi C_A$), there is a 
continuous curve formed by the pairs ($C,\alpha_s$), for which one finds physical solutions. Indeed, for small values of $C$ one
has that no solution exists;  this absence of  solutions persists (for the quenched case) until the  critical value $\overline{C}\approx0.56$  is reached, after which  one   finds  exactly  one   monotonically  decreasing  solution.  However,  for values  up  to  $C\approx0.8$ the  coupling needed  to get  the corresponding  running mass  is of  ${\cal O}(1)$, while for the quenched case  the expected coupling from the 4-loop (momentum subtraction) calculation   is  $\alpha_s=0.22$   at  $\mu=4.3$   GeV~\cite{Boucaud:2005rm}.  This latter value is obtained for $C\approx1.8$ -- $1.9$, whereas for $C\approx0.88$ one finds the solution to~\1eq{me-gen} for the 
lowest order perturbative value of the coefficient. In general one observes, as expected, that as $C$ is increased, $\alpha_s$ decreases, \eg for   $C=1.1$, $1.3$, $1.5$ and $1.7$  one  obtains solutions  corresponding  to  the strong coupling values $\alpha_s\approx0.53$, $0.39$, $0.30$, and $0.25$, respectively.  

In the right panel of~\fig{masseq} we plot the solutions for the most representative $C$ values, \ie $C=0.88$ and $C=1.85$ (corresponding to, as already said,  $\alpha_s\approx0.88$ and $0.22$ respectively), normalized in such a way that the mass at zero coincides with the IR saturating value found in lattice (Landau gauge) quenched simulations~\cite{Bogolubsky:2009dc}, or $m^2(0)=\Delta^{-1}(0)\approx0.141$ GeV$^2$. 
As can be readily appreciated, the masses obtained 
display the basic qualitative features expected on  
general field-theoretic considerations and employed in numerous phenomenological  studies; 
in particular, they are monotonically decreasing functions of the momentum, 
and vanish rather rapidly in the ultraviolet~\cite{Cornwall:1981zr,Lavelle:1991ve,Aguilar:2007ie}. 
It would seem, therefore, that the PT-BFM all-order analysis described here puts 
the entire concept of the gluon mass, and a variety of fundamental properties ascribed to it, on a   
solid first-principle basis.


\end{document}